\PassOptionsToPackage{dvipsnames}{xcolor}
\PassOptionsToPackage{table}{xcolor}
\documentclass[sigconf, nonacm]{acmart}

\usepackage{xspace}
\usepackage{listings}
\usepackage{subcaption}
\usepackage{hyperref}
\PassOptionsToPackage{hyphens}{url}
\usepackage{url}
\usepackage{framed}
\usepackage{xcolor}
\usepackage{multirow}
\usepackage{pifont}
\usepackage{enumitem}

\newcommand{\jaralyzer}{\textsc{Jaralyzer}\xspace}

\newcommand{\unshade}{\textsc{Unshade}\xspace}

\definecolor{lightgray}{gray}{0.9}
\newcommand{\llbox}[1]{%
	\vspace{4pt}%
	\noindent%
	\setlength{\fboxsep}{6pt}%
	\fcolorbox{black}{lightgray}{%
		\parbox{0.93\linewidth}{#1}%
	}%
}

\setcopyright{none}

\begin{document}

	\title{Uncovering Hidden Inclusions of Vulnerable Dependencies in Real-World Java Projects}

	\author{Stefan Schott}
	\affiliation{%
		\institution{Paderborn University}
		\city{Paderborn}
		\country{Germany}
	}
	\email{stefan.schott@upb.de}
	\orcid{0000-0002-0644-3297}
	
	\author{Serena Elisa Ponta}
	\affiliation{%
		\institution{SAP Labs}
		\city{Mougins}
		\country{France}
	}
	\email{serena.ponta@sap.com}
	\orcid{0000-0002-6208-4743}
	
	\author{Wolfram Fischer}
	\affiliation{%
		\institution{SAP SE}
		\city{Stuttgart}
		\country{Germany}
	}
	\email{wolfram.fischer@sap.com}
	\orcid{0000-0001-8127-8837}
	
	\author{Jonas Klauke}
	\affiliation{%
		\institution{Paderborn University}
		\city{Paderborn}
		\country{Germany}
	}
	\email{jonas.klauke@upb.de}
	\orcid{0000-0001-9160-9636}
	
	\author{Eric Bodden}
	\affiliation{%
		\institution{Paderborn University \& \\ Fraunhofer IEM}
		\city{Paderborn}
		\country{Germany}
	} 
	\email{eric.bodden@upb.de}
	\orcid{0000-0003-3470-3647}
	
	\begin{abstract}
		Open-source software (OSS) dependencies are a dominant component of modern software code bases.
Using proven and well-tested OSS components lets developers reduce development time and cost while improving quality.
However, heavy reliance on open-source software also introduces significant security risks, including the incorporation of known vulnerabilities into the codebase.
To mitigate these risks, metadata-based dependency scanners, which are lightweight and fast, and code-centric scanners, which enable the detection of modified dependencies hidden from metadata-based approaches, have been developed.

In this paper, we present \unshade, a \emph{hybrid} approach towards dependency scanning in Java that combines the efficiency of metadata-based scanning with the ability to detect modified dependencies of code-centric approaches.
\unshade first \emph{augments} a Java project’s software bill of materials (SBOM) by identifying modified and hidden dependencies via a bytecode-based fingerprinting mechanism.
This augmented SBOM is then passed to a metadata-based vulnerability scanner to identify known vulnerabilities in both declared and newly revealed dependencies.

Leveraging \unshade’s high scalability, we conducted a large-scale study of the 1,808 most popular open-source Java Maven projects on GitHub. 
The results show that nearly 50\% of these projects contain at least one modified, hidden dependency associated with a known vulnerability. 
On average, each affected project includes more than eight such hidden vulnerable dependencies, all missed by traditional metadata-based scanners. 
Overall, \unshade identified 7,712 unique CVEs in hidden dependencies that would remain undetected when relying on metadata-based scanning alone.
	\end{abstract}

	
	\maketitle

	\section{Introduction}
\label{sec:introduction}

The use of open-source software (OSS) has become integral to modern software development. 
Today, OSS is so widespread that third-party components often make up the largest share of a project’s total code base. 
A 2023 Endor Labs report on dependency management~\cite{stateOfDependencyManagement2023} found that, on average, 71\% of the code in typical Java projects originates from OSS.
This number continues to rise as an increasing number of OSS components find their way into modern software projects~\cite{openlogicReport2025}.
Consequently, vulnerable open-source dependencies represent a significant threat to modern software systems, which is also reflected in the OWASP Top Ten list of critical application security risks~\cite{owaspTopTen}. 
High-profile incidents such as \textit{log4shell}~\cite{everson2022log4shell} and the \textit{Equifax breach}~\cite{luszcz2018apache}, both of which had a profound impact on the cybersecurity landscape, further highlight the dangers posed by insecure OSS components.

To remedy this risk, various open-source~\cite{owaspDC, osvScanner, retireJS, ponta2020detection, schottJaralyzer} and commercial \textit{dependency scanners}~\cite{snykScanner, blackduckScanner, mendScanner, endorlabsScanner} have been developed, which seek to identify inclusions of vulnerable OSS dependencies within software projects.
According to Schott et al.~\cite{schottJaralyzer}, these dependency scanners can generally be categorized into two main types: \textit{metadata-based} and \textit{code-centric} scanners.

Metadata-based scanners analyze project dependencies by relying on the metadata associated with each inclusion. 
A common approach is to generate a Software Bill of Materials (SBOM) for the project under inspection. 
The SBOM enumerates all dependencies used in the software, along with their identifiers and versions. 
This list is then matched against a security advisory cataloging known vulnerable dependency versions, commonly expressed as Common Vulnerability Enumeration (CVE) entries~\cite{cve}.

Code-centric scanners, on the other hand, analyze the actual code of included dependencies to determine whether they contain known vulnerabilities.
These scanners rely on a knowledge base of \emph{fix commits}, which represent the actual code changes made to OSS projects to patch specific vulnerabilities.
By comparing the code of included dependencies against these fix commits, code-centric scanners can identify whether a dependency contains vulnerable code that has not been patched yet.

Both approaches have distinct advantages and disadvantages.
Metadata-based scanners are generally faster and easier to implement, as they do not require deep code analysis.
However, they cannot reliably identify vulnerable dependencies when modifications have been applied to them, such as \emph{re-bundling} or \emph{re-packaging}.
According to Dann et al~\cite{dann2021identifying}, and Dietrich et al.~\cite{dietrich2023security}, these modifications are widespread in the Java ecosystem and can even be found in the Java standard library.
Such modifications typically \emph{hide} the inclusion of a vulnerable dependency.
Additionally, they usually alter the metadata of the dependency, rendering it unrecognizable to metadata-based scanners.
Code-centric scanners, in contrast, can identify inclusions of vulnerable dependencies even in modified form, as they analyze their actual code.
However, due to the complexity of code analysis, they are generally slower and more resource-intensive.
In addition, these scanners require fix commits to be available, which is not the case for all vulnerabilities, and assume that such commits can be compiled. 
However, as Schott et al.~\cite{schottJaralyzer} show, successful compilation is not always feasible.
These limitations make it difficult for code-centric scanners to be applied on a large scale.

To combat these issues, we propose \unshade, a \emph{hybrid} approach to Java dependency scanning that combines the strengths of both metadata-based and code-centric scanners.
\unshade first creates a knowledge base of \emph{fingerprints} for all known-to-be-vulnerable OSS versions from a given security advisory.
This fingerprint represents the actual code that comprises the vulnerable dependency.
During scan time, \unshade uses the SBOM of the project under inspection to identify included dependencies and their versions.
For each listed dependency, \unshade retrieves its corresponding artifact and tries to identify whether any fingerprint from the knowledge base is present in the dependency's bytecode, indicating a modified dependency inclusion.
If this is the case, \unshade \emph{augments} the original SBOM with the newly identified, but hidden, vulnerable dependencies.
This augmented SBOM is then passed to a metadata-based scanner to report the final list of vulnerabilities.

Leveraging the high scalability of \unshade, we conducted a large-scale study and applied it on the 1,808 most popular Java Maven projects on GitHub, with each more than 500 stars.
We identified 899 (49.7\%) projects that included at least one modified known-to-be-vulnerable dependency.
In total \unshade detected 7,680 re-bundled or re-packaged known-to-be-vulnerable dependencies across these projects, leading to the identification of 7,712 CVE entries that were previously not reported by the metadata-based OSV scanner.
These results highlight the prevalence of modified vulnerable dependencies in real-world Java projects and demonstrate the effectiveness of \unshade in uncovering these hidden vulnerabilities.

This paper makes the following contributions:

\begin{itemize}
    \item We propose \unshade, a hybrid approach to Java dependency scanning that combines the strengths of metadata-based and code-centric scanners.
    \item We present a large-scale study demonstrating the prevalence of modified vulnerable dependencies in real-world Java projects.
    \item We provide an open-source implementation of \unshade~\footnote{\url{https://github.com/stschott/unshade}}.
\end{itemize}

	\section{Background}
\label{sec:background}

In the following we introduce the fundamental topics of how Java dependencies are included and modified.
Furthermore, we we explain \emph{fully qualified names} (FQN) in Java and how they are altered when dependencies are re-packaged.

\subsection{Java Dependencies}
\label{subsec:dependencyInclusion}

The term \emph{dependency} denotes a software library or framework that is distributed independently and integrated into a software project \cite{dann2021identifying}. 
In the Java ecosystem, such libraries are commonly packaged as JAR archives that contain the library's bytecode. 
Bytecode is a machine-executable intermediate representation produced by compiling Java source code, enabling execution on the Java Virtual Machine (JVM).

The management of dependencies is typically handled through build automation tools such as Maven or Gradle. 
When using Maven, developers declare the required library by specifying its \texttt{groupId}, \texttt{artifactId}, and \texttt{version} within one or multiple \texttt{pom.xml} files, which define the project's build configuration. 
Together, these three attributes, commonly referred to as \emph{GAV}, uniquely identify a dependency.

\subsection{Dependency Modifications}
\label{subsec:dependecyModifications}

In some cases, developers modify dependencies in Java projects, which are then incorporated into other projects in their altered form.
Dann et al.~\cite{dann2021identifying} identified and classified these modifications into four distinct types.

\noindent
\textbf{Type 1 (patched):}
\label{subsubsec:type1}
This type of modification often arises when developers fork the source code of an OSS project to apply changes or patches. 
Dependencies altered in this manner do not contain the original bytecode but instead include the bytecode resulting from re-compiling the modified source code. Additionally, the GAV is typically adjusted by appending a suffix such as \texttt{fix} or \texttt{patch} to indicate the modification.

\noindent
\textbf{Type 2 (Uber-JAR):}
\label{subsubsec:type2}
In this type of modification, developers \emph{re-bundle} multiple OSS components into a single JAR file, commonly referred to as an Uber-JAR. 
This modification is sometimes signaled by the JAR file name, which may include terms such as \texttt{jar-with-dependencies} or \texttt{uber}. 
Unlike type 1, the original bytecode of each individual OSS component is preserved.

\noindent
\textbf{Type 3 (bare Uber-JAR):}
\label{subsubsec:type3}
Type 3 modifications are similar to type 2 in that multiple OSS components are re-bundled into a single JAR file. 
However, in contrast to type 2, the metadata contained within the JAR such as the \texttt{pom.xml} files, the \texttt{META-INF} directory and file timestamps are removed. 
This type of modification is most commonly observed in legacy JAR files that were created prior to the introduction of assembly plugins.

\noindent
\textbf{Type 4 (re-packaged Uber-JAR):}
\label{subsubsec:type4}
Type 4 modifications are also related to type 2. 
However, rather than just re-bundling multiple OSS components into a single JAR file, the OSS are \emph{re-packaged}. 
This process modifies the original package names, either by adding a prefix or by replacing them entirely with new names \cite{mavenShadePluginRepackaging}. 
Such modifications are commonly applied to prevent naming collisions and version conflicts. 
As a consequence, the original bytecode of the OSS is substantially altered, since all internal references within the code must be updated accordingly.

In their analysis of a sample of 254 vulnerable classes, Dann et al. identified 67,196 artifacts on Maven Central that contained these classes in modified form, underscoring the widespread occurrence of such modifications.
They also found that types 2 and 3 were the most prevalent modification types, accounting for over 90\% of all identified modified artifacts \cite{dann2021identifying}.

In the remainder of this paper, we focus on identifying vulnerable dependencies that have been modified as described in types 2, 3, and 4.
We refer to modifications of types 2 and 3 collectively as \emph{re-bundling} and to type 4 modifications as \emph{re-packaging}.

\subsection{Fully Qualified Name (FQN)}
\label{subsec:fqn}

In Java, each class is defined within a specific package that serves as a namespace to organize and group related classes together. 
The \emph{fully qualified name} (FQN) of a class uniquely identifies it by combining its package name with the class name. 
For example, consider a class named \texttt{Foo} located in the package \texttt{com.example.utils}. 
The FQN of this class would be \texttt{com.example.utils.Foo}.

When a dependency is re-packaged (type 4 modification), the original package names are altered, which in turn changes the FQNs of all classes within that dependency.
For instance, if the package \texttt{com.example.utils} is re-packaged to \texttt{org.modified.utils}, the FQN of \texttt{Foo} would change to \texttt{org.modified.utils.Foo}.
This change of FQNs affects not only class identifiers but also field and method signatures, as these likewise incorporate the fully qualified names of their types and declaring classes.
Furthermore, each reference within the bytecode that points to the original FQN must also be updated to reflect the new package structure, which significantly impacts the bytecode of the modified dependency.
This includes references in method calls, field accesses, type casts, exception handling, etc.

In the remainder of this paper, the term \emph{unqualified names} denotes class names as well as method and field signatures that do not include any package information.
	\section{\unshade}
\label{sec:concept}

In the following we present \unshade, a \emph{hybrid} approach towards identifying known-to-be-vulnerable Java dependencies.

\subsection{Overview of \unshade}
\label{sec:overview}

\begin{figure}
	\includegraphics[width=\linewidth]{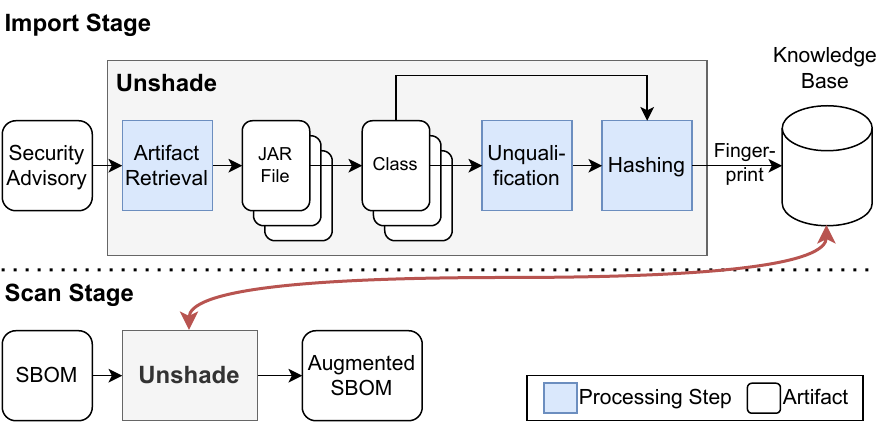}
	\caption{Overview of \unshade}
	\label{fig:overview}
\end{figure}

Figure~\ref{fig:overview} illustrates the overall workflow of \unshade.
The approach behind \unshade can be divided into two separate stages: the \emph{import} stage and the \emph{scan} stage.

The import stage is only executed once to create a knowledge base that \unshade uses during the scan stage for each project scan.
In the import stage, a security advisory such as the OSV database~\cite{osvAdvisory} serves as input.
\unshade extract the identifiers of \emph{all} artifacts associated with a vulnerability from the advisory.
These artifact identifiers are then used to retrieve the corresponding \texttt{JAR} files from a public artifact repository such as Maven Central.
Then \unshade extracts all classes from the respective JAR files which are processed one by one.
First, \unshade computes a \emph{hash} value that uniquely represents the class's bytecode structure.
However, just using the hash value calculated from the raw bytecode is not sufficient, as re-packagings modify the bytecode and thus change the hash value.
Therefore, \unshade additionally applies an \emph{unqualification} step that removes all package-related information from the bytecode before computing a second hash value.
Both hash values, the \emph{qualified} and the \emph{unqualified} hash, serve as a \emph{fingerprint} for each class and are stored in a database together with the information about what class and artifact they belong to.

In the scan stage, \unshade takes the SBOM of a project as input.
From the SBOM, \unshade retrieves all declared dependencies and downloads their corresponding JAR files from a public artifact repository.
Then, similar to the import stage, \unshade extracts all classes from the JAR files and computes both the qualified and unqualified hash values for each class.
Now, to identify potentially re-bundled or re-packaged vulnerable dependencies, \unshade checks if any of the stored fingerprints in the knowledge base are contained in the scanned JAR files.
Finally, \unshade enriches the SBOM with the identified vulnerable dependencies and outputs an \emph{augmented} SBOM that can be analyzed by a given dependency scanner.

\subsection{Import Stage}
\label{sec:import}

During the \emph{import} stage, \unshade creates a knowledge base of known-to-be-vulnerable artifacts that is used during the scan stage.
This is a one-time effort that needs to be extended only when new security advisories are published.
As input, \unshade takes a security advisory such as the OSV database~\cite{osvAdvisory}.
From the advisory, \unshade extracts the identifiers of \emph{all} artifacts associated with a vulnerability.
Each entry in the advisory maps a vulnerability identifier, typically a CVE ID, to one or more artifact identifiers, typically Maven coordinates (group ID, artifact ID, version).
\unshade uses these artifact identifiers to retrieve the corresponding \texttt{JAR} files from a public artifact repository such as Maven Central.
Then, \unshade extracts all classes from the respective JAR files which are processed one by one and generates a fingerprint that uniquely represents the artifact.
After processing all artifacts, \unshade stores the generated fingerprints in a database together with the information about what artifact they belong to.

\subsubsection{Hashing}
\label{sec:hashing}

To generate a fingerprint for each artifact, \unshade computes two hash values for each class within that artifact: a \emph{qualified} hash and an \emph{unqualified} hash.
The qualified hash is computed directly from the raw bytecode of each class.
To do so, \unshade uses the XXHash3 algorithm~\cite{xxhash} to calculate a 128-bit hash value from the bytecode.
For our use case of identifying re-bundled artifacts, there is no need for cryptographic security.
However, speed and collision resistance are extremely important as \unshade needs to process potentially millions of classes during the import stage.
XXHash3 provides a good trade-off between these requirements.

For detecting re-bundled artifacts, this hash is sufficient as the bytecode remains unchanged when an artifact is re-bundled without modifications (see Section~\ref{subsec:dependecyModifications}).
However, when an artifact is re-packaged, the package structure of its classes is altered. This changes their fully qualified names (FQNs), which propagates to the bytecode and ultimately results in a different hash value (see Section~\ref{subsec:fqn}).
Here, simply computing the hash from the raw bytecode is not sufficient.

\subsubsection{Unqualification}
\label{sec:unqualification}

To address the issue of altered bytecode in re-packaged artifacts, \unshade needs to create a hash value that is independent of changes to FQNs.
To achieve this, \unshade applies an \emph{unqualification} step that removes all package-related information from the bytecode before computing the hash value.
This process involves identifying and replacing all occurrences of FQNs in the bytecode with their corresponding unqualified names.
This includes class identifiers, method signatures, field signatures, and all references within the bytecode that point to the corresponding FQNs.
By removing package-related information, the resulting bytecode reflects only the structural aspects of the class and omits details that change as a result of re-packaging.

The typical way of how re-packagings are applied is by using the Maven Shade Plugin~\cite{mavenShadePluginRepackaging}.
The plugin uses the ASM library~\cite{asmLibrary} to perform this transformation.
\unshade leverages the same mechanism to perform the unqualification step.

\subsubsection{Fingerprint}
\label{sec:fingerprint}

Once both the qualified and unqualified hash values are computed for each class within an artifact, \unshade aggregates these hash values to create a fingerprint for the entire artifact.
A fingerprint $F$ for each artifact is defined as follows:
\[
F = (A, Q, U)
\]
where $A$ is the unique artifact identifier (groupId, artifactId, version), $Q$ is the set of qualified hashes of all classes within the artifact, and $U$ is the set of unqualified hashes of all classes within the artifact.
This fingerprint is then stored in \unshade's knowledge base to be used during the scan stage.
It can be used to uniquely identify the presence of the artifact, even if it has been re-bundled or re-packaged.

\subsection{Scan Stage}
\label{sec:scan}

In the \emph{scan} stage, \unshade analyzes all dependencies used in a given software project to identify re-bundled or re-packaged inclusions of known-to-be-vulnerable artifacts.
As input, \unshade takes a Software Bill of Materials (SBOM) that lists all dependencies used in the project along with their corresponding artifact identifiers (group ID, artifact ID, version).
This SBOM typically comes in the form of a \texttt{CycloneDX}~\cite{cycloneDX} or \texttt{SPDX}~\cite{spdx} document that can automatically be generated by build tools such as Maven or Gradle using appropriate plugins.
As in the import stage, for each dependency listed in the SBOM, \unshade retrieves the corresponding artifact, extracts all classes from the respective JAR file, and computes both the qualified and unqualified hash values for each class.
Doing so, \unshade generates the set $Q_{scan}$ of qualified hashes and the set $U_{scan}$ of unqualified hashes for all classes within the dependency being analyzed.
Next, \unshade compares these sets against all fingerprints stored in its knowledge base to identify potential matches with known-to-be-vulnerable artifacts.
A match can be of two types: a \emph{qualified match} or an \emph{unqualified match}.
A \emph{qualified match} occurs when \unshade finds a subset relation between the set of qualified hashes of the dependency being analyzed ($Q_{scan}$) and the set of qualified hashes of a known-to-be-vulnerable artifact ($Q_{KB}$) belonging to fingerprint $F_{KB}$ within the knowledge base:
\[
Q_{KB} \subseteq Q_{scan}
\]
Such a match indicates that the known-to-be-vulnerable artifact has been re-bundled within the analyzed dependency and is thus included in the software project.

Analogously, an \emph{unqualified match} occurs when \unshade finds a subset relation between the set of unqualified hashes of the dependency being analyzed ($U_{scan}$) and the set of unqualified hashes of a known-to-be-vulnerable artifact ($U_{KB}$) belonging to fingerprint $F_{KB}$ within the knowledge base:
\[
U_{KB} \subseteq U_{scan}
\]
Such a match indicates that the known-to-be-vulnerable artifact has been re-packaged within the analyzed dependency.

\unshade will repeat this process for all dependencies listed in the provided SBOM.
At the end of the scan stage, \unshade will \emph{augment} the original SBOM by adding information about any identified re-bundled or re-packaged known-to-be-vulnerable artifacts.
This augmented SBOM can then be used by dependency scanners, which themselves are not capable of detecting re-bundled or re-packaged inclusions.
	\newcommand{\RQONE}{What is the prevalence of modified dependency inclusions in real-world Java projects?}
\newcommand{\RQTWO}{How can \unshade help uncover hidden vulnerabilities in these modified dependencies that are missed by traditional metadata-based vulnerability scanners?}
\newcommand{\RQTHREE}{What is the performance overhead of using \unshade in a vulnerability scanning workflow?}

\section{Study}
\label{sec:evaluation}

Using \unshade, we conducted a large-scale study across 1,808 popular Java Maven projects on GitHub.
We aimed to identify the prevalence of \emph{modified} known-to-be-vulnerable dependencies in real-world projects and evaluate the effectiveness of \unshade in uncovering these hidden vulnerabilities.
We answer the following research questions.

\begin{enumerate}[rightmargin=0.9em]
	\item[\textbf{RQ1:}] \RQONE
	\item[\textbf{RQ2:}] \RQTWO
	\item[\textbf{RQ3:}] \RQTHREE
\end{enumerate}

The first research question (RQ1) focuses on understanding how common it is for Java projects to include \emph{modified} versions of dependencies that are known to be vulnerable and gives a unique insight into the prevalence of such dependencies.
The second research question (RQ2) evaluates the effectiveness of \unshade in identifying hidden vulnerabilities within these modified dependencies that traditional metadata-based vulnerability scanners may overlook.
Finally, the third research question (RQ3) assesses the performance overhead introduced by integrating \unshade into a typical vulnerability scanning workflow, providing insights into its practicality for real-world applications.

\subsection{Study Setup}
\label{sec:setup}

\begin{figure}
	\includegraphics[width=\linewidth]{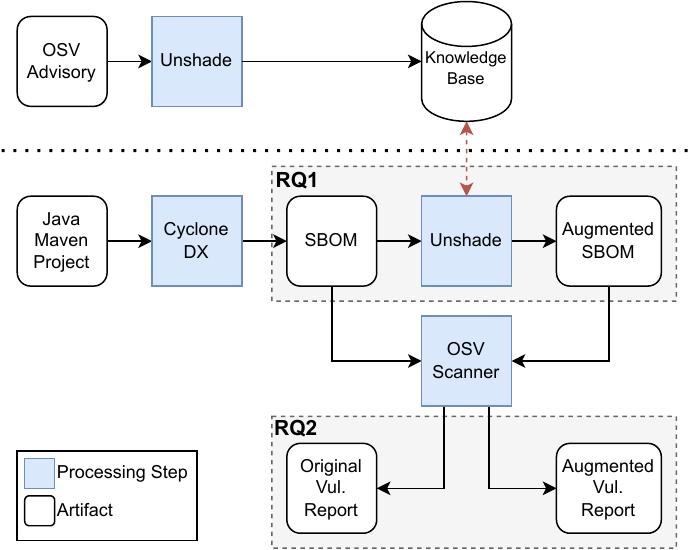}
	\caption{Overview of the study setup}
	\label{fig:evaluationOverview}
\end{figure}

Figure~\ref{fig:evaluationOverview} depicts an overview of our study setup.
First we started with creating \unshade's database of known-to-be-vulnerable dependencies.
To do so, we provided Google's OSV Advisory~\cite{osvAdvisory} to \unshade's import stage to process and fingerprint all listed vulnerable dependency artifacts.
As of our import cutoff date (2025-08-12), OSV contained 5,557 unique Maven vulnerability entries.
We collected all dependency artifacts referenced by these entries and retrieved all the ones hosted on Maven Central.
In total, we were able to retrieve 94,903 unique dependency artifacts from Maven Central.
We did not retrieve all artifacts referenced by OSV as some of them, such as Jenkins plugins, are not hosted on Maven Central.
These 94,903 artifacts contained a total of 81,689,087 Java classes (an average of 860 classes per artifact) that we then fingerprinted and stored in \unshade's knowledge base.
This knowledge base served as the foundation for the remainder of our study.

To select popular real-world Java project for our study, we used the SEART~\cite{dabic2021sampling} dataset.
SEART is a large-scale dataset of open-source software projects on GitHub that was specifically designed for MSR (mining software repositories) studies.
First, we filtered the SEART dataset to include only projects whose primary language is Java and that have more than 500 stars on GitHub.
After this filtering step, we obtained 6,864 Java projects hosted on GitHub.
Next, we further filtered these projects to include only those that use Maven as their build system.
To do so, we cloned every project and checked for the presence of a \texttt{pom.xml} file in the root directory.
After this step, we were left with 2,294 Java Maven projects.
To each of these projects, we applied the CycloneDX Maven Plugin~\cite{cycloneDxPlugin} to generate an SBOM for each project.
This SBOM contains the complete list of dependencies used by the project, including their unique artifact identifiers.
Finally, we excluded projects for which the SBOM generation failed or timed out after 15 minutes, resulting in a final dataset of 1,808 Java Maven projects for our study.
We provided each generated SBOM to \unshade, which analyzed each project for modified, known-to-be-vulnerable dependencies and augmented the SBOMs with any identified re-bundled or re-packaged dependencies that were not present in the original SBOM.

After augmenting the SBOMs with \unshade, we passed both the original and augmented SBOMs to Google's metadata-based OSV scanner~\cite{osvScanner} to identify inclusions of known-to-be-vulnerable dependencies within the provided SBOMs.
Finally, we compared the vulnerability reports generated from the original and augmented SBOMs to identify vulnerable dependencies newly revealed by the augmented SBOMs that were not detected by the scanner when only the original SBOM was provided.

All experiments were conducted in a Docker container (v29.1.3) running Alpine Linux (v3.19) configured with Eclipse Temurin JDK 17 and Maven 3.9.6.
The docker container was running on a machine using four cores of an Intel Xeon E5-2695 v3 (2.30GHz) CPU with 32GB of RAM.
Furthermore, we used v2.9.1 of the CycloneDX Maven Plugin for SBOM generation and v2.2.2 of the OSV scanner.
We used the default configuration for the CycloneDX Maven Plugin, which by default excludes test dependencies from the generated SBOMs.

\subsection{RQ1: \RQONE}
\label{sec:rq1}

In this section, we present the findings of our investigation into the prevalence of modified dependency inclusions in real-world Java projects.
In this investigation, we specifically focused on identifying modified dependencies that are relevant in a security context, i.e., dependencies that are known to be vulnerable according to the OSV security advisory.
This means that each of the identified modified dependencies corresponds to a known vulnerability.
As described in our study setup (Section~\ref{sec:setup}), we analyzed a dataset of the most popular 1,808 Java Maven projects hosted on GitHub.
We generated an SBOM for each project using CycloneDX, augmented it with \unshade and compared the original SBOM to its augmented counterpart.

\begin{table}[]
	\caption{Summary of RQ1 results}
	\centering
		\begin{tabular}{l|r}
        Projects                        & 1,808                     \\ \hline
        Projects with Modified Dependencies     & 899                       \\ \hline
        Bundling Artifacts              & 3,641                     \\ \hline
        Bundled Artifacts               & 19,052                    \\ \hline
        Bundled Artifacts (grouped)     & 7,680                     \\ \hline
        Re-bundled Artifacts (grouped)  & 5,847                     \\ \hline
        Re-packaged Artifacts (grouped) & 1,833                    
        \end{tabular}
	\label{tab:rq1_results}
\end{table}

Table~\ref{tab:rq1_results} shows the summary of our findings for RQ1.
In total, \unshade identified in 899 of the 1,808 analyzed projects at least one modified known-to-be-vulnerable dependency inclusion not present in the original SBOM.
This means that nearly 50\% of the analyzed projects included at least one modified known-to-be-vulnerable dependency, highlighting the prevalence of such modified dependency inclusions in real-world projects.
These dependency inclusions are systematically missed by state-of-the-art SBOM generators such as CycloneDX and, as shown in multiple independent studies~\cite{schottJaralyzer,dann2021identifying,dietrich2023security}, remain undetected by current dependency scanners.

Amongst these 899 projects with modified dependencies, we found a total of 3,641 artifacts that either re-bundled or re-packaged at least one other known-to-be-vulnerable dependency.
Most of these artifacts include multiple modified dependencies.
In total, we identified 19,052 modified dependency inclusions not present in the original SBOMs.
However, in many cases multiple versions of the same artifact contain the exact same bytecode.
This is typically the case when a project consists of multiple modules.
When a new version of such a multi-module project is released, each module is assigned a new version number, even though the bytecode of some modules remains unchanged.
Accounting for this, we grouped all identified inclusions of different versions of the same artifact that share the same bytecode into a single entry.
After this grouping step, we identified a total of 7,680 unique modified dependency inclusions across all analyzed projects.
Out of these 7,680 unique modified dependency inclusions, 5,847 (76.1\%) were re-bundled dependencies, while the remaining 1,833 (23.9\%) were re-packaged dependencies.

\begin{table}[]
	\caption{Distribution of the most commonly bundled artifacts}
	\centering
		\begin{tabular}{l|r|r}
        \textbf{Artifact}   & \textbf{Count} & \textbf{Share} \\ \hline
        netty-codec-http    & 584            & 7.6\%          \\
        tomcat-coyote       & 532            & 6.9\%          \\
        tomcat-catalina     & 528            & 6.9\%          \\
        tomcat-util         & 480            & 6.3\%          \\
        tomcat-websocket    & 397            & 5.2\%          \\
        netty-codec-haproxy & 343            & 4.5\%          \\
        commons-lang3       & 282            & 3.7\%          \\
        log4j-1.2.17        & 260            & 3.4\%          \\
        netty-common        & 252            & 3.3\%          \\
        bcprov-jdk15to18    & 251            & 3.3\%          \\
        Others (181)        & 3,771          & 48.9\%        
        \end{tabular}
	\label{tab:rq1_distribution}
\end{table}

Table~\ref{tab:rq1_distribution} shows the distribution of the most commonly bundled artifacts across all analyzed projects.
In total, we identified 191 unique artifacts (ignoring different versions) that were bundled as modified dependencies in at least one project.
However, only ten of these artifacts account for nearly 51.1\% of all identified modified dependency inclusions, mostly being netty and tomcat modules.
This indicates that certain libraries are more prone to being re-bundled or re-packaged by developers, potentially due to their widespread use or specific functionality.

In summary, our findings for RQ1 demonstrate that modified dependency inclusions are prevalent in real-world Java projects, with nearly half of the analyzed projects including at least one modified dependency, with most projects even containing multiple such inclusions.
It further demonstrates that the state-of-the-art SBOM generation tool CycloneDX systematically misses these modified dependencies, creating knowledge gaps in the SBOM.
It is important to note that our study focused specifically on modified dependencies that are known to be vulnerable.
Thus, we only considered the 94,903 artifacts listed in the OSV security advisory and exclusively identified re-bundlings and re-packagings of these artifacts.  
These vulnerabilities are not necessarily exploitable since usually only parts of a dependency are actually used by the including project.
However, the unknown presence of such artifacts in a project still poses a security risk, especially when the vulnerability is present in a part of the dependency that is actually used.
Therefore, our findings only present a lower bound on the actual prevalence of modified dependency inclusions in real-world Java projects.
Considering other artifacts beyond those listed in OSV would likely reveal an even higher prevalence of modified dependencies.
In the future this study can be extended to investigate e.g. licensing issues.

\llbox{
    Nearly 50\% of the 1,808 most popular Java Maven projects on GitHub contain at least one modified dependency inclusion linked to a vulnerability that is not detected by the CycloneDX SBOM generation tool.
    On average, each of these projects includes more than eight re-bundled or re-packaged artifacts.
}
\section{RQ2: \RQTWO}
\label{sec:rq2}

Multiple independent studies \cite{schottJaralyzer,dann2021identifying,dietrich2023security} have already shown that state-of-the-art metadata-based vulnerability scanners often fail to identify dependencies that are known to be vulnerable when those dependencies have been modified by the including project, for example through re-bundling or re-packaging.
However, these studies have not investigated the extent of this problem in real-world projects, instead focusing on small-scale benchmarks.
In this section, we investigate the extent to which state-of-the-art dependency scanners miss known-to-be-vulnerable dependencies in real-world Java projects due to modifications made by the including project.
Furthermore, we evaluate the effectiveness of augmenting the original SBOMs with \unshade in uncovering these previously hidden vulnerabilities.

As described in our study setup (Section~\ref{sec:setup}), we analyzed a dataset of the most popular 1,808 Java Maven projects hosted on GitHub.
For this investigation, we used the original SBOMs and the augmented SBOMs generated in RQ1.
We provided both the original and augmented SBOMs to Google's metadata-based OSV scanner~\cite{osvScanner} to identify inclusions of known-to-be-vulnerable dependencies within the provided SBOMs.
Finally, we compared the vulnerability reports generated from the original and augmented SBOMs to identify vulnerable dependencies newly revealed by the augmented SBOMs that were not detected by the scanner when only the original SBOM was provided.
Note that although the OSV scanner also reports vulnerabilities without assigned CVE identifiers, we restrict our analysis to CVE-based vulnerabilities. 
Consequently, our results represent a lower bound on the number of newly detected known-to-be-vulnerable dependencies.

Since our knowledge base is based on the OSV security advisory, all re-bundled or re-packaged dependencies identified by \unshade correspond to at least one known vulnerability.
Thus, any modified known-to-be-vulnerable dependency inclusion identified in RQ1 that is not present in the original SBOM will lead to at least one newly detected vulnerability when the augmented SBOM is used.
In total, across all 7,680 identified bundled artifacts (grouped), the OSV scanner reported 102,823 CVE entries associated with these artifacts. 
A single CVE entry is often associated with multiple artifacts, as vulnerabilities may affect multiple versions of a dependency or multiple modules within a project. 
After removing duplicate CVE identifiers across all bundled artifacts, as well as CVE identifiers that were also reported by the OSV scanner for original, non-modified dependencies, we identified 7,712 unique CVE entries that would have been missed when relying solely on the original SBOMs and were revealed only through the augmented SBOMs.

To ensure that the missing CVE entries are not solely attributable to modified artifacts being absent from the original SBOMs, we empirically verified, using a random sample, whether applying the OSV scanner directly to the projects (allowing it to generate the SBOM internally) would reveal additional artifact inclusions or CVE entries. 
This was not the case, confirming that the OSV scanner itself does not identify modified dependencies that are not present in the provided SBOM

These results further confirm the findings of previous studies \cite{schottJaralyzer,dann2021identifying,dietrich2023security} that state-of-the-art metadata-based vulnerability scanners often miss known-to-be-vulnerable dependencies when those dependencies have been modified by the including project.
Moreover, our results demonstrate that this is not just a problem in synthetic benchmarks, but a prevalent issue in real-world Java projects on a large scale.
By augmenting the original SBOMs with \unshade, we were able to leverage already existing tools like the OSV scanner to detect vulnerabilities that would otherwise remain hidden.

\llbox{
    Across the 899 projects containing modified dependencies, we uncovered 7,712 unique CVE entries that the OSV scanner would have missed when relying solely on the original SBOMs.
    This shows that augmenting SBOMs uncovers many otherwise hidden vulnerabilities, highlighting the real-world impact of modifications.
}
\section{RQ3: \RQTHREE}
\label{sec:rq3}

In the previous sections, we demonstrated both the prevalence of modified, known-to-be-vulnerable dependencies in popular real-world Java projects and the blind spots of state-of-the-art metadata-based vulnerability scanners in detecting them.
Furthermore, we showed that \unshade can help uncover these hidden vulnerabilities by augmenting the original SBOMs with modified dependencies, enabling existing scanners to detect vulnerabilities that would otherwise remain unseen.
However, integrating \unshade into a typical vulnerability scanning workflow introduces additional computational overhead.
In this section, we evaluate the performance overhead of using \unshade in a vulnerability scanning workflow.

\begin{table}[]
	\caption{Runtime of different stages in the vulnerability scanning workflow}
	\centering
		\begin{tabular}{l|r|r|r}
                & \textbf{Average} & \textbf{Median} & \textbf{Maximum} \\ \hline
        SBOM Generation & 18.2s           & 6.6s           & 505.0s          \\
        Unshade         & 16.4s           & 13.6s          & 148.6s          \\
        SBOM Scan       & 3.2s            & 2.9s           & 36.6s           \\
        Aug. SBOM Scan  & 3.6s            & 3.2s           & 29.3s          
        \end{tabular}
	\label{tab:rq3_results}
\end{table}

Table~\ref{tab:rq3_results} summarizes the runtime of the different stages in the vulnerability scanning workflow for the 1,808 analyzed Java Maven projects.
The SBOM generation stage, which uses the CycloneDX Maven Plugin to generate an SBOM for each project, has an average runtime of 18.2 seconds per project, with a median of 6.6 seconds and a maximum of 505.0 seconds.
Note that the measured runtime does not include the time needed to download the corresponding artifacts from remote repositories, such as Maven Central, as we already cached all required artifacts locally before starting the experiments.
This caching step ensures that network latency does not influence the measured runtimes.
The \unshade stage, which analyzes each project for modified, known-to-be-vulnerable dependencies and augments the SBOMs accordingly, has an average runtime of 16.4 seconds per project, with a median of 13.6 seconds and a maximum of 148.6 seconds.
Finally, the SBOM scanning stage, which uses Google's OSV scanner to identify known-to-be-vulnerable dependencies within the provided SBOMs, has an average runtime of 3.2 seconds per project for the original SBOMs and 3.6 seconds per project for the augmented SBOMs. 
Since we directly provide the SBOMs to the OSV scanner, the measured runtimes do not include the time needed to generate the SBOMs internally, which would otherwise be required if the scanner were applied directly to the projects.
These results indicate that integrating \unshade into a typical vulnerability scanning workflow introduces only moderate performance overhead, with most time spent on SBOM generation, which is already required in standard scanning workflows.
Overall, using \unshade to augment the SBOM and scan it with a metadata-based dependency scanner is considerably faster than employing a code-centric scanner~\cite{schottJaralyzer}.

\llbox{
    Integrating \unshade into a vulnerability scanning workflow introduces only a moderate performance overhead with an average of 16.4 seconds per project scan. 
}
	\section{Threats to Validity}
\label{sec:threats}

There are a few potential threats to the validity of our study that need to be considered when interpreting our results.
First, our study focuses exclusively on Java projects that use Maven as their build system.
While Maven is a widely used build system in the Java ecosystem, there are other build systems such as Gradle and Ant that are also popular.
Furthermore, the vulnerable dependencies we uncovered using \unshade must not necessarily be exploitable in the context of the including project.
Typically only parts of a dependency are actually used by the including project, and if the vulnerable parts are not used, the vulnerability may not be exploitable.
However, the unknown presence of such artifacts in a project still poses a potential security risk that needs to be assessed.
Another potential threat to validity is our reliance on the OSV security advisory as the basis of our study.
While OSV is a comprehensive and widely used security advisory, it may not cover all known vulnerabilities in the Java ecosystem.
Furthermore, as shown by Schott et al.~\cite{schottJaralyzer}, different advisories may disagree on which artifacts are affected by a given vulnerability.
Thus, our results may differ if we had used a different security advisory or combined multiple advisories.
Finally, since our study is based on the OSV security advisory, we focused our evaluation of \unshade on the OSV scanner.
While OSV is a state-of-the-art metadata-based vulnerability scanner, there are other scanners available that may have different detection capabilities.

	\section{Related Work}
\label{sec:relatedWork}

Legislation, such as the EU Cyber Resilience Act~\cite{euCRA} and the US Cybersecurity Executive Order~\cite{usCeo}, mandates software development companies to ensure the secure use of open-source software (OSS) components.
Furthermore, they require the provision of software bills of materials (SBOMs) for software products that list each component used in the software.
To comply with these regulations, two primary SBOM standards have emerged.
The Software Package Data Exchange (SPDX)~\cite{spdx} format, created by the Linux Foundation, is a widely adopted open standard for SBOMs that provides a comprehensive framework for documenting software components, their licenses, and associated metadata.
Another standard is the CycloneDX~\cite{cycloneDX} standard, created by the OWASP Foundation, which focuses on security and supply chain management, providing a lightweight format for representing software components and their relationships.
To identify inclusions of vulnerable OSS dependencies in software projects, various dependency scanners and software composition analysis (SCA) tools have been created.
Popular open-source tools include OWASP Dependency-Check~\cite{owaspDC} and Google's OSV Scanner~\cite{osvScanner}.
Another frequently used free-to-use dependency scanner is GitHub Dependabot~\cite{dependabot}, which is typically seamlessly integrated into development workflows.
Commercial SCA tools include Endor Labs SCA~\cite{endorlabsScanner}, Snyk Open Source SCA~\cite{snykScanner}, Mend SCA~\cite{mendScanner}, Black Duck SCA~\cite{blackduckScanner} and SonarQube~\cite{sonarqube}.
These tools are metadata-based and primarily rely on metadata files to identify dependencies.
Eclipse Steady~\cite{ponta2020detection, ponta2018beyond, plate2015impact} is a code-centric approach that goes beyond metadata-based approaches by analyzing the actual code to identify vulnerable dependencies more accurately.

Dann et al.~\cite{dann2021identifying} conducted a study towards the prevalence of modified OSS dependencies in Java projects.
Overall, they identified four separate types of modifications that developers apply to dependencies.
Furthermore, they evaluated the impact of these modifications on state-of-the-art dependency scanners and found that such modifications significantly decrease the accuracy of these scanners.
Dietrich et al.~\cite{dietrich2023security} further investigate the impact of re-bundled dependencies on the accuracy of dependency scanners.
Furthermore, they present a lightweight approach to identify re-bundled dependencies in Java applications.
Schott et al.~\cite{schottJaralyzer} present \jaralyzer, a bytecode-centric dependency scanner for Java applications that can accurately identify inclusions of vulnerable dependencies, even if they are modified.
In a comparison against five state-of-the-art metadata-based and code-centric dependency scanners, \jaralyzer was the only tool that was consistently able to identify inclusions of vulnerable dependencies.
Zhao et al.~\cite{zhao2023software} conducted an empirical study on the effectiveness of SCA tools for vulnerability detection in Java projects.
They evaluated six tools and found that all tools exhibited inaccuracies.
Wu et al.~\cite{wu2025more} evaluated the effectiveness of six SBOM tools in accurately generating SBOMs for Java applications.
They found that the detection capabilities of these tools vary significantly, especially when considering different import scenarios.
	\section{Conclusion}
\label{sec:conclusion}

This paper presents \unshade, a \emph{hybrid} approach towards dependency scanning in Java that combines the efficiency of metadata-based scanning with the ability to detect modified dependencies of code-centric approaches.
\unshade first \emph{augments} a Java project’s software bill of materials (SBOM) by identifying modified dependencies via a bytecode-based fingerprinting mechanism.
Doing so allows \unshade to identify re-bundled and re-packaged dependencies that are normally hidden to traditional tools.
This augmented SBOM is then passed to a metadata-based vulnerability scanner to identify known vulnerabilities in both declared and newly revealed dependencies.

Leveraging \unshade’s high scalability, we conducted a large-scale study of the 1,808 most popular open-source Java Maven projects on GitHub. 
The results show that nearly 50\% of these projects contain at least one modified dependency associated with a known vulnerability. 
On average, each affected project includes more than eight such hidden vulnerable dependencies, all missed by traditional metadata-based scanners.
Overall, \unshade identified 7,712 unique CVEs in hidden dependencies that would remain undetected when relying on metadata-based scanning alone.

These findings highlight the high prevalence of modified dependencies in real-world Java projects and the significant security risks they pose.
By combining the strengths of metadata-based and code-centric approaches, \unshade provides a practical and effective solution for enhancing dependency scanning in Java projects.

	\bibliographystyle{ACM-Reference-Format}
	\bibliography{literature}

@inproceedings{schottJaralyzer,
	author       = {Stefan Schott and
	Serena Elisa Ponta and
	Wolfram Fischer and
	Jonas Klauke and
	Eric Bodden},
	title        = {Bytecode-centric Detection of Known-to-be-vulnerable Dependencies in Java Projects},
	booktitle = {2026 IEEE/ACM 48th International Conference on Software Engineering (ICSE)},
	year = {2026},
}

@article{ponta2020detection,
	title={Detection, assessment and mitigation of vulnerabilities in open source dependencies},
	author={Ponta, Serena Elisa and Plate, Henrik and Sabetta, Antonino},
	journal={Empirical Software Engineering},
	volume={25},
	number={5},
	pages={3175--3215},
	year={2020},
	publisher={Springer}
}

@inproceedings{ponta2018beyond,
	title={Beyond metadata: Code-centric and usage-based analysis of known vulnerabilities in open-source software},
	author={Ponta, Serena Elisa and Plate, Henrik and Sabetta, Antonino},
	booktitle={2018 IEEE International Conference on Software Maintenance and Evolution (ICSME)},
	pages={449--460},
	year={2018},
	organization={IEEE}
}

@article{dann2021identifying,
	title={Identifying challenges for oss vulnerability scanners-a study \& test suite},
	author={Dann, Andreas and Plate, Henrik and Hermann, Ben and Ponta, Serena Elisa and Bodden, Eric},
	journal={IEEE Transactions on Software Engineering},
	volume={48},
	number={9},
	pages={3613--3625},
	year={2021},
	publisher={IEEE}
}

@inproceedings{dietrich2023security,
	title={On the security blind spots of software composition analysis},
	author={Dietrich, Jens and Rasheed, Shawn and Jordan, Alexander and White, Tim},
	booktitle={Proceedings of the 2024 Workshop on Software Supply Chain Offensive Research and Ecosystem Defenses},
	pages={77--87},
	year={2024}
}

@misc{openlogicReport2025,
	author={Perforce Software Inc.},
	title={{2025 State of Open Source Report}},
	url={https://www.openlogic.com/resources/state-of-open-source-report},
	note={{Accessed} 2024-12-13},
	year={2023}
}

@misc{stateOfDependencyManagement2023,
	author={Plate, Henrik},
	title={{State of Dependency Management 2023}},
	url={https://www.endorlabs.com/learn/state-of-dependency-management-2023},
	note={{Accessed} 2025-12-12},
	year={2025}
}

@article{luszcz2018apache,
	title={Apache struts 2: how technical and development gaps caused the equifax breach},
	author={Luszcz, Jeff},
	journal={Network Security},
	volume={2018},
	number={1},
	pages={5--8},
	year={2018},
	publisher={Elsevier}
}

@inproceedings{everson2022log4shell,
	title={Log4shell: Redefining the web attack surface},
	author={Everson, Douglas and Cheng, Long and Zhang, Zhenkai},
	booktitle={Proc. Workshop Meas., Attacks, Defenses Web (MADWeb)},
	pages={1--8},
	year={2022}
}

@misc{owaspTopTen,
	author={OWASP Foundation},
	title={{OWASP Top Ten}},
	url={https://owasp.org/www-project-top-ten/},
	note={{Accessed} 2024-12-13},
	year={2021}
}

@misc{owaspDC,
	author={OWASP},
	title={{OWASP Dependency-Check}},
	url={https://owasp.org/www-project-dependency-check/},
	note={{Accessed} 2024-12-13},
	year={2024}
}

@misc{dependabot,
	author={GitHub Inc.},
	title={{GitHub Dependabot Alerts}},
	url={https://docs.github.com/en/code-security/dependabot/dependabot-alerts/about-dependabot-alerts},
	note={{Accessed} 2024-12-13},
	year={2024}
}

@misc{osvScanner,
	author={Google Inc.},
	title={{OSV Scanner}},
	url={https://google.github.io/osv-scanner/},
	note={{Accessed} 2024-12-13},
	year={2024}
}

@misc{retireJS,
	author={RertireJS},
	title={{Retire.js}},
	url={https://retirejs.github.io/retire.js/},
	note={{Accessed} 2024-12-16},
	year={2024}
}

@misc{snykScanner,
	author={Snyk Limited},
	title={{Snyk Open Source SCA}},
	url={https://snyk.io/product/open-source-security-management/},
	note={{Accessed} 2024-12-13},
	year={2024}
}

@misc{blackduckScanner,
	author={Black Duck Software Inc},
	title={{Black Duck SCA}},
	url={https://www.blackduck.com/software-composition-analysis-tools/black-duck-sca.html},
	note={{Accessed} 2024-12-13},
	year={2024}
}

@misc{mendScanner,
	author={Mend.io},
	title={{Mend SCA}},
	url={https://www.mend.io/sca/},
	note={{Accessed} 2024-12-13},
	year={2024}
}

@misc{endorlabsScanner,
	author={Endor Labs},
	title={{Endor Labs SCA}},
	url={https://www.endorlabs.com/use-cases/reachability-based-sca},
	note={{Accessed} 2024-12-13},
	year={2024}
}

@misc{cve,
	author={The MITRE Corporation},
	title={{About the CVE Program}},
	url={https://www.cve.org/About/Overview},
	note={{Accessed} 2024-12-13},
	year={2024}
}

@inproceedings{plate2015impact,
	title={Impact assessment for vulnerabilities in open-source software libraries},
	author={Plate, Henrik and Ponta, Serena Elisa and Sabetta, Antonino},
	booktitle={2015 IEEE International Conference on Software Maintenance and Evolution (ICSME)},
	pages={411--420},
	year={2015},
	organization={IEEE}
}

@misc{mavenShadePluginRepackaging,
	author={The Apache Software Foundation},
	title={{Apache Maven Shade Plugin - Relocating Classes}},
	url={https://maven.apache.org/plugins/maven-shade-plugin/examples/class-relocation.html},
	note={{Accessed} 2025-06-23},
	year={2025}
}

@misc{osvAdvisory,
	author={Google Inc.},
	title={{OSV - A distributed vulnerability database for Open Source}},
	url={https://osv.dev/},
	note={{Accessed} 2025-12-15},
	year={2025}
}

@misc{xxhash,
	author={Cyan4973},
	title={{xxHash - Extremely fast hash algorithm}},
	url={https://github.com/Cyan4973/xxHash},
	note={{Accessed} 2025-12-16},
	year={2025}
}

@misc{asmLibrary,
	author={OW2 Consortium},
	title={{ASM - A Java bytecode manipulation and analysis framework}},
	url={https://asm.ow2.io/},
	note={{Accessed} 2025-12-16},
	year={2025}
}

@misc{cycloneDX,
	author={OWASP Foundation},
	title={{CycloneDX - A Software Bill of Materials (SBOM) standard}},
	url={https://cyclonedx.org/},
	note={{Accessed} 2025-12-17},
	year={2025}
}

@misc{cycloneDxPlugin,
	author={OWASP Foundation},
	title={{CycloneDX Maven Plugin}},
	url={https://github.com/CycloneDX/cyclonedx-maven-plugin},
	note={{Accessed} 2025-12-19},
	year={2025}
}

@misc{spdx,
	author={The Linux Foundation},
	title={{SPDX - The System Package Data Exchange}},
	url={https://spdx.dev/},
	note={{Accessed} 2025-12-17},
	year={2025}
}

@misc{euCRA,
	author={European Union},
	title={{EU Cyber Resilience Act}},
	url={https://digital-strategy.ec.europa.eu/en/policies/cyber-resilience-act},
	note={{Accessed} 2026-01-15},
	year={2024}
}

@misc{usCeo,
	author={The White House},
	title={{US Cybersecurity Executive Order}},
	url={https://www.federalregister.gov/documents/2021/05/17/2021-10460/improving-the-nations-cybersecurity},
	note={{Accessed} 2026-01-15},
	year={2021}
}

@inproceedings{dabic2021sampling,
  title={Sampling projects in github for MSR studies},
  author={Dabic, Ozren and Aghajani, Emad and Bavota, Gabriele},
  booktitle={2021 IEEE/ACM 18th International Conference on Mining Software Repositories (MSR)},
  pages={560--564},
  year={2021},
  organization={IEEE}
}

@misc{sonarqube,
	author={SonarSource Sarl},
	title={{SonarQube}},
	url={https://www.sonarqube.org/},
	note={{Accessed} 2026-01-15},
	year={2026}
}

@inproceedings{zhao2023software,
  title={Software composition analysis for vulnerability detection: An empirical study on Java projects},
  author={Zhao, Lida and Chen, Sen and Xu, Zhengzi and Liu, Chengwei and Zhang, Lyuye and Wu, Jiahui and Sun, Jun and Liu, Yang},
  booktitle={Proceedings of the 31st ACM Joint European Software Engineering Conference and Symposium on the Foundations of Software Engineering},
  pages={960--972},
  year={2023}
}

@article{wu2025more,
  title={More Than Meets the Eye: On Evaluating SBOM Tools In Java},
  author={Wu, Menghan and Zhao, Yukai and Hu, Xing and Zhan, Xian and Li, Shanping and Xia, Xin},
  journal={ACM Transactions on Software Engineering and Methodology},
  year={2025},
  publisher={ACM New York, NY}
}
\end{document}